\documentclass[runningheads]{llncs}
\usepackage{booktabs}
\usepackage{array}
\usepackage{amsfonts}
\usepackage{nicefrac}
\usepackage{microtype}
\usepackage{subcaption}
\usepackage{caption}
\usepackage{adjustbox}
\usepackage{xcolor}
\usepackage{xspace}
\usepackage[T1]{fontenc}
\usepackage{graphicx}
\newcommand{\mypara}[1]{
\vspace{6pt}
\noindent
\textbf{#1}.
}

\newcommand{\mylist}[1]{
\vspace{4pt}
\noindent
\textbf{#1}.
}

\newcommand{\bench}{\texttt{RepoVBench}\xspace}
\newcommand{\verusbench}{\texttt{VerusBench}\xspace}
\newcommand{\verismo}{\texttt{VeriSMo}\xspace}
\newcommand{\ragverus}{\textsc{RagVerus}\xspace}
\newcommand{\autoverus}{\textsc{AutoVerus}\xspace}

\begin{document}
\title{\ragverus: Repository-Level Program Verification with LLMs using Retrieval Augmented Generation}
\titlerunning{Retrieval Augmented Generation Verifying Rust}
\author{Si Cheng Zhong\inst{1} \and
Jiading Zhu\inst{1}\textsuperscript{$\dagger$} \and
Yifang Tian\inst{1}\textsuperscript{$\dagger$} \and
Xujie Si\inst{1}}
\authorrunning{S. Zhong et al.}
%
\institute{University of Toronto, Canada \\
\email{sicheng.zhong@mail.utoronto.ca}, \email{six@cs.toronto.edu}
}
\maketitle              
\begin{abstract}

Scaling automated formal verification to real-world projects requires resolving cross-module dependencies and global contexts, which are challenges overlooked by existing function-centric methods. We introduce \ragverus, a framework that synergizes retrieval-augmented generation with context-aware prompting to automate proof synthesis for multi-module repositories, achieving a \textit{27\%} relative improvement on our novel \bench benchmark—the first repository-level dataset for Verus with \texttt{383} proof completion tasks. \ragverus \textit{triples} proof pass rates on existing benchmarks under constrained language model budgets, demonstrating a scalable and sample-efficient verification. 

\end{abstract}
\section{Introduction}

Good engineers favor well-written tests to confirm that code works correctly, at least for the tested inputs. In software verification, testing is replaced by a set of complete specification; a verifier conducts a compile-time check to ensure the implementation matches the specification, providing vital assurances especially for high-confidence systems in critical areas such as operating systems and financial applications~\cite{li2024survey,zhang2024selene}. However, proof construction is a challenging process that requires domain expertise~\cite{wen2024enchanting,lattuada2024verusExp,li2023veriScale}. This process is time-consuming and non-trivial especially for large software projects.

Advances in generative AI for code completion have spurred interest in verification-aware program synthesis, aiming to simultaneously enhance code trustworthiness and automate proof generation~\cite{kamath2023loopy,yang2024autoverus,sun2024clover}. Large language models (LLMs) streamline formal verification by rapidly sampling and iterating proof candidates, lowering barriers to adoption~\cite{chen2024evolveVerus,aggarwal2024alphaverus,chakraborty2024FStar}. However, current function-level program verification approaches inherit vanilla retrieval models~\cite{zhang2024selene,chakraborty2024FStar} from theorem proving that solely rely on finetuned embedding encoders using prior examples~\cite{yang2023leandojo}, lacking the rich, precise domain knowledge of code syntax or project structure~\cite{shrivastava2023repoPrompt}. Moreover, existing datasets~\cite{misu2024MbppDfy} are derived for small-scale, isolated contexts, lacking evaluation in complex repository settings to reflect real practices in production.

We recognize the main challenges of repository-level verification tasks to be (1) the vast semantic context and (2) the identification of correct dependent premises, as noted in prior works studying large-scale verification in the theorem proving domain~\cite{li2024survey,yang2023leandojo,klein2014sel4}.
A major challenge in LLM proof generation is premise retrieval—selecting the minimal lemma set from a repository-level codebase that aids verification while fitting within the context window. Scaling verification to repository level adds complexity due to inter-dependencies in large codebases, while the lack of benchmark data limits furthur evaluatiosn in the area.

To assess LLM-based formal repository-level program verification, we establish three main contributions in this paper:

We propose \ragverus, a new verification framework integrating retrieval-augmented generation (RAG) with context-aware prompting to guide LLMs in synthesizing verifiable code for interdependent, large-scale codebases. We provide a configurable RAG pipeline to enable systematic evaluation of diverse retrieval strategies (e.g., semantic search, dependency graphs) within a sandboxed Verus environment, informing cross-module dependencies and project-specific invariants.

We also construct \bench, the first repository-level program verification benchmark, where we collect recent award-winning Verus projects and extract  programs with complex dependencies. 
Specifically,
we build a dataset containing 2073 functions with dependencies over 52 modules, supporting evaluation on 383 proof completion tasks. 

We further evaluate \ragverus on two benchmarks: \verusbench from \autoverus ~\cite{yang2024autoverus}, which is a function-level verification benchmark, and the new repository-level verification benchmark, \bench.
On \verusbench, \ragverus demonstrates a 200+\% improvement in proof pass rates compared to the non-RAG baseline under constrained sampling budgets.
On \bench, our framework produces contextually coherent proofs to solve 5\% more of the benchmark (27\% relative increase), outperforming isolated function-level approaches.

\section{Background}

\textbf{Verus}~\cite{lattuada2023verus} is an SMT-based verification tool designed to formally verify Rust programs while leveraging Rust's powerful type system, in particular its linearity and borrow checking mechanisms. Users write its domain-specific language (DSL) inline with Rust, while dividing Rust code into three code modes: \textit{specification} annotation, \textit{proof} annotation, and \textit{executable} implementation.
Verification in Verus is performed by encoding Rust functions and their associated annotations as SMT formulas~\cite{cho2024framework}, which are then processed by an SMT solver such as Z3~\cite{de2008z3}.
Recently, \autoverus~\cite{yang2024autoverus}, an LLM-based approach shows strong performance in automating proof completion in Verus solving nearly all tasks of VerusBench.

\textbf{RAG}~\cite{NEURIPS2020_6b493230} is a novel technique to enhance generative models by incorporating external knowledge retrieval, enabling more accurate and contextually relevant responses in knowledge-intensive tasks.
Existing RAG methods~\cite{misu2024MbppDfy,zhang2024selene} used in the software verification domain are relatively simple, such as directly comparing the method signatures and post-conditions. We draw inspiration from repository-level code generation~\cite{shrivastava2023repoPrompt} and theorem proving~\cite{yang2023leandojo} domains, examining a framework to capture the structured nature of programming tasks.

\section{\ragverus Framework}

\vspace*{-\baselineskip}
\begin{figure}[h!]
    \centering
    \centerline{
        \includegraphics[width=1.1\textwidth]{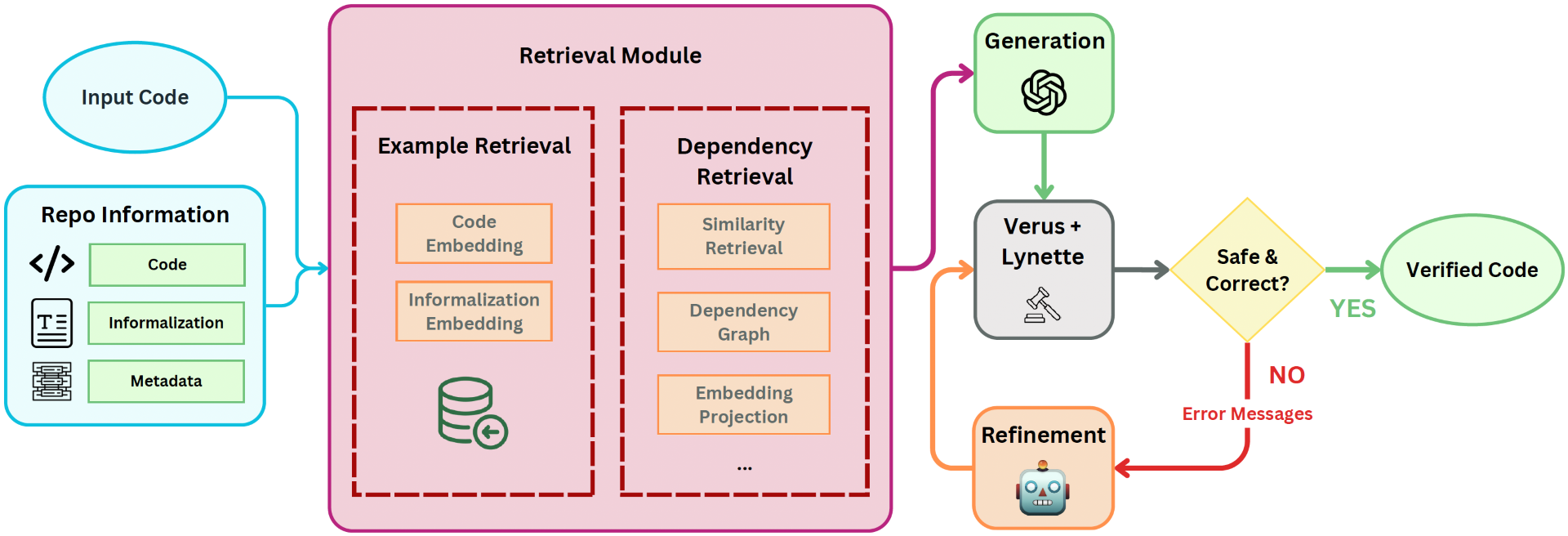}
    }
    \caption{Illustration of the \ragverus Framework Pipeline} 
    \label{fig:RAG_Verus}
\end{figure}
\vspace*{-\baselineskip}

We propose \ragverus, a retrieval-augmented framework for LLM-based program verification, with a special focus on complex Verus projects. The framework modularizes verification task preparation, context retrieval methods, and proof generation models, leveraging repository information and verifier feedback to align domain knowledge.

\subsection{The General Pipeline}
As illustrated in Fig.~\ref{fig:RAG_Verus}, the \ragverus pipeline consists of three stages: 1) mining code properties, 2) retrieving task-specific information, and 3) finally generating proofs and/or verifiable code.

\mypara{Mining Code Properties}
We preprocess repository-wide code artifacts to extract verification-critical metadata (more details in Section 4.1), building on insights from context modeling in repository-level code generation~\cite{shrivastava2023repoPrompt}. Through static analysis 
over the codebase hierarchy,
we identifying function/type signatures, method calls, module dependencies, and control/dataflow relationships essential for proof construction. 

To enable efficient retrieval, \ragverus uses persistent vector stores where documents are encoded into embeddings using OpenAI's \texttt{text-embedding-3-large} \cite{openai2024embedding}, indexed via FAISS \cite{faisspaper}. Depending on the retrieval method chosen, either the Rust code themselves or their metadata is indexed. This hybrid representation preserves both syntactic and semantic patterns in the form of high-dimensional embeddings, enabling adaptive context retrieval in later stages.

\mypara{Context Retrieval}
We believe two general kinds of retrieval tasks are essential for repository verification. \textbf{\textit{Few-shot example retrieval}} matches code snippets or metadata to provide contextual proof patterns and exemplar code-proof pairs, while \textbf{\textit{Dependency retrieval}} identifies function signatures and domain-specific syntax necessary for constructing proofs. Retrieved contexts are integrated into prompts to guide proof generation, with examples expanded in Section~\ref{sec:instantiations}.

\mypara{Proof Generation}
This stage accepts a generic code generation agent like \autoverus.
The framework processes Rust modules containing executable code and Verus specifications (pre-/post-conditions) \textit{without} existing proof annotations. Language models synthesize verification annotations—loop invariants, assertions, and proof blocks—conditioned on retrieved contexts to produce a fully annotated Rust program. Verus compiler validates if specifications hold across all situations, with failed proofs triggering iterative refinement—feeding back compiler errors to the agent—if requested. 

\vspace{3pt}
While being evaluated on proof generation tasks, \ragverus can also support flexible extensions for code generation as well as specification inference.

\subsection{Instantiations of Retrieval Module}
\label{sec:instantiations}

Building upon the modular framework established in Section 3.1, this subsection presents customizable retrieval modules tailored to address distinct program verification challenges, 
each designed to align with specific task requirements within the verification pipeline.

\subsubsection{Few-Shot Example Retrieval}

Few-shot example retrieval in \ragverus involves selecting a small number of representative examples to guide the LLM's automated proof generation. The objective is to retrieve relevant code snippets or contexts as contextual prompts, ensuring that generated annotations adhere to correct styles and practices. To achieve this, the retrieval process leverages FAISS~\cite{faisspaper} to identify the most semantically similar examples, implemented through LlamaIndex~\cite{llamaindex}. During retrieval, depending on the method chosen, either the input code or its metadata is used as the query vector. The FAISS vector store then retrieves the top-k most similar examples, with an ID filter ensuring that the input document itself is excluded from the retrieved examples. 

Two separate vector indices are created for each dataset: one for code body embeddings and one for code informalization embeddings. The embeddings are built only with unverified files, but during retrieval, the corresponding verified files are also retrieved, forming input-output pairs as few-shot examples. We retain a maximum of three examples after ranking, prioritizing diversity and relevance. These examples provide contextual references that enhance the LLM's ability to generate relevant and accurate proofs~\cite{brown2020fewshot}, ultimately improving the overall quality and precision of the verification process.

\begin{itemize}
    \item
\mylist{Code-Based Example Selection}
Searching by code embeddings effectively provides relevant reference examples by capturing structural and functional similarities between code implementations. This is particularly beneficial for the task of program verification, as it ensures that retrieved examples align with correct coding patterns and verification logic, improving the accuracy and consistency of proof generation.
    \item
\mylist{Informalization-Based Example Selection}
This process generates natural language contexts of the target Rust function as supplementary information for retrieval. It takes the code implementation and the essential metadata~\cite{shrivastava2023repoPrompt} of each Rust function as input and produces a detailed natural language description of the function behaviour, which we refer to as the informalized code summary. Matching these summaries by semantic similarities during context retrieval enhances knowledge of relevant topics and verification flow on the topic.
\end{itemize}

\vspace*{-1.8\baselineskip}
\begin{figure}[h!]
    \setlength{\abovecaptionskip}{0pt}
    \setlength{\belowcaptionskip}{-10pt}
    \centering
    \includegraphics[width=0.8\textwidth]{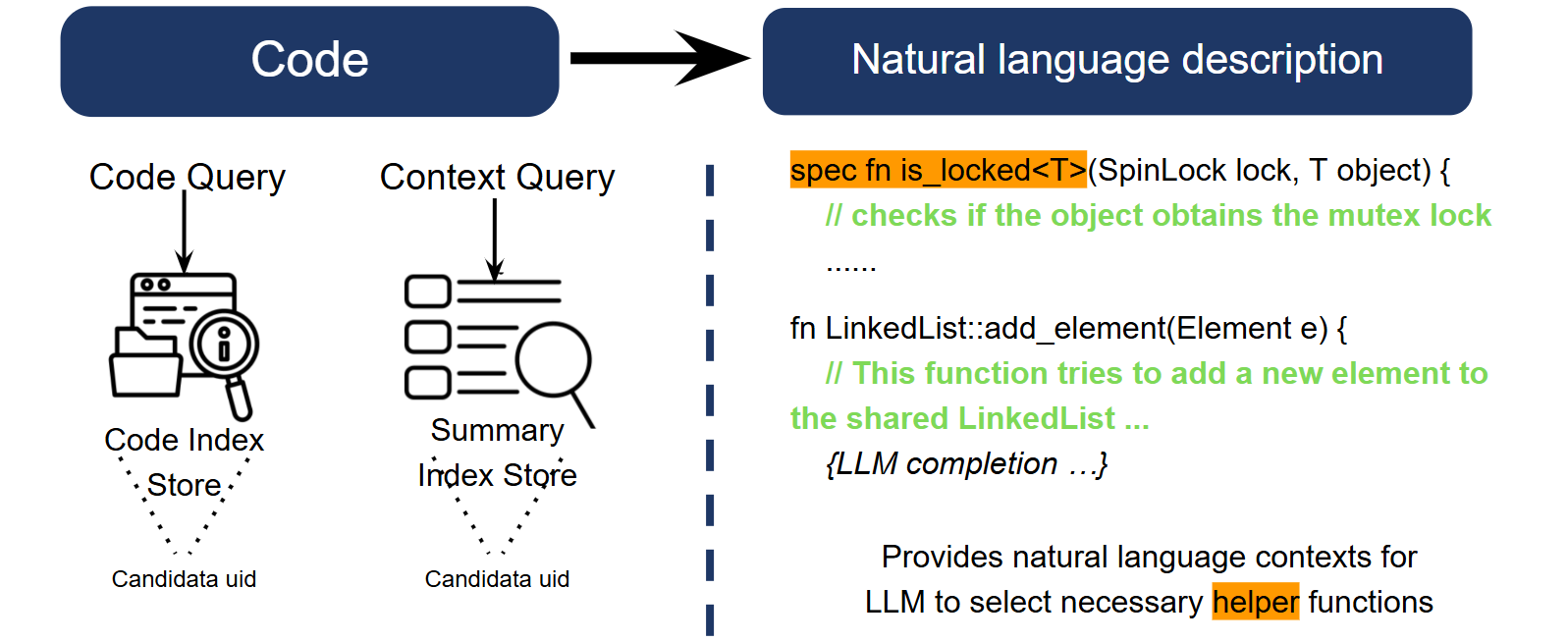}
    \caption{Illustration of code informalization utility.} 
    \label{fig:code_informalization}
\end{figure}
\vspace*{-1.5\baselineskip}

\subsubsection{Dependency Retrieval}
Dependency retrieval identifies all relevant function signatures within the code repository that should serve as dependent premises for generating proofs for a particular task. A well-chosen premise pool provides essential context, such as file dependencies and variable typing, ensuring accurate and complete proof generation. Understanding premise dependencies involves knowing available functions and their interactions within the codebase. Accurately identifying these relevant functions creates an essential toolkit for the LLM to elaborate on, streamlining the generation process and minimizing redundant generation. We recommend three primary methods:

\begin{itemize}
    \item
\mylist{Embedding-based Dependency Retrieval}  
This method heuristically operates with the assumption that functions similar to the input function can serve as its premises; for example, a function \texttt{less\_than} can serve as premise for \texttt{less\_equal}. Similarity is determined using embeddings derived from both the code itself and its functional summary. FAISS is used to perform nearest-neighbor retrieval in the embedding space.
    \item
\mylist{Finetuned Projection Matching}  
In this approach, a specialized embedding model is trained to project the embedding of the input code close to the embeddings of its logical premises. The model is trained using a set of human-labeled function-premise pairs, allowing it to learn the relationships between input functions and their relevant premises.
    \item
\mylist{Dependency Graph}  
This method leverages static analysis on the abstract syntax tree generated by the compiler to construct a dependency graph for the codebase. The premises of the input function are derived by traversing the call hierarchies and extracting functions on which the input function depends directly or indirectly.
\end{itemize}

\noindent 
These methods complement each other, with similarity retrieval providing a heuristic baseline, embedding projection leveraging learned relationships, and the dependency graph offering a deterministic, compiler-supported solution.

\section{\bench: A Repository-level Verification Benchmark}

Existing benchmarks focus on function-level proof generation~\cite{yang2024autoverus,aggarwal2024alphaverus}, where all contexts are self-contained in single files, and have become saturated—with tools like \autoverus achieving over 90\% success rates—offering little challenge or differentiation for emerging techniques. We introduce \textbf{\bench} to address the absence of repository-level benchmarks in Verus,
capturing complexities of full-scale codebases involving inter-module interactions and environmental dependencies.

Our new benchmark consists of three core components: \textit{Code} (a curated set of repositories utilizing Verus), \textit{Environment} (a standardized build setup), and \textit{Metric} (evaluating verification effectiveness). We provide users with flexible control over running experiments, enabling customized assessments that reflect practical challenges in formal verification.

\subsection{Creating Benchmark from Code Repositories}

We generate benchmark datasets from Verus repositories by processing the Rust files through two modules that flexibly \textit{create downstream tasks} and \textit{build supporting contexts for retrieval agents}.

\mypara{Code Masking with Verus Property Extractor}
We build a new extraction tool utilizing a property parser from the Verus compiler to identify formal properties such as preconditions, postconditions, and invariants. We leverage the \textit{code mode} hierarchy to precisely control over \texttt{exec}, \texttt{spec}, and \texttt{proof} code, enabling the flexible creation of various downstream tasks for evaluation. For instance, to create a proof completion task, we erase proof lines from the verified functions or modules to form queries. See Appendix \ref{appen:bench_example} for an illustrative example.

\mypara{Metadata Generation}
To build supporting contexts for the benchmark, we extract key metadata from Rust functions. We create a unique index of content by breaking down constructs like type and class definitions, making each function searchable for retrieval tasks.
The metadata contains 1) various names including the file name, function name and construct name, 2) function's type signature, 3) methods invocations, 4) type identifiers and variable declarations appearing in the function, and 5) the function's code mode. We use code mode to assess whether a function can serve as potential dependencies in a selected type of task, guided by the \texttt{exec-proof-spec} hierarchy's constraint that higher-assurance modes (\texttt{proof}) cannot invoke lower-assurance implementations (\texttt{exec}).

\subsection{\bench Data Source and Environment}

We build the initial \bench upon the \verismo project~\cite{zhou2024verismo}, a verified security module developed for confidential virtual machines (VMs) running on a specific AMD architecture. Using Rust and the Verus verification tool, \verismo performs permission-based reasoning to ensure memory safety and confidentiality.

We extracted 2,073 code pieces from \verismo, including macro definitions and functions, and parsed 1,656 indexable functions,
out of which we identified 460 functions containing \texttt{proof} lines in the original code using the syntax tracer. We further filtered out functions that the compiler can automatically solve even without proof code, which resulted in \textbf{383} tasks for the benchmark. These tasks are functions that include \texttt{spec} in their definitions and require \texttt{proof} bodies to be completed, which in total consists of \textbf{8,108} lines of proof, a mean of 17 lines of proof per verification task (with a median of 8). 
We analyze the ground truth proofs and split the benchmark into two categories based on whether the proofs depend on other function calls ---  52 \textbf{Simple} tasks, which have no dependencies, and 331 \textbf{Complex} tasks, which require dependencies. 

\mypara{Environment}
We develop a build pipeline to evaluate \bench in an encapsulated environment that ensures reproducible verification outcomes and enhances ease of experiment. For every request to test some varifiable code, we make a segregated work tree to track code changes, compile independently, and revert to clean states, allowing concurrent execution of different experiments.

\vspace{5pt}
\noindent
\textbf{Ongoing and Future Effort}.
We are in the process of expanding \bench by adding a wider array of verification repositories.
One immediate goal is to include Anvil~\cite{sun2024anvil}, a formal tool built upon Verus for verifying Kubernetes controller correctness.
We anticipate \bench will keep growing as Verus becomes increasingly popular over time.

\subsection{Metric}
The primary metric for evaluating the verification dataset is the number of \textbf{success} - tasks proven to be both correct and safe: 

\begin{itemize}
    \item 
\textbf{Correctness}
The code is correct if the generated annotations pass Verus compilation, which checks for constraint propagation throughout the project, validating the required specifications and logical correctness.
    \item 
\textbf{Safety}
The code is safe\footnote{following terminology in \autoverus~\cite{yang2024autoverus} referring untampered implementation and specification as \textit{safe} code} if the implementation lines stays unaltered, thus maintaining the original functionalities. We utilize the Lynette checker from \autoverus to examine the code consistency.
\end{itemize}

As verification success is a rigid binary metric, the BLEU score similarity is more suitable for signaling gradual improvement. We may also consider the \textit{style} of the generated code by comparing it to the ground-truth reference using BLEU score to assess the alignment with expected coding standards.

\section{Evaluation}

We run the evaluation pipeline against two datasets respectively: \verusbench ~\cite{yang2024autoverus} and \bench that we created. Although \verusbench is not a repository-level verification dataset, we conducted this experiment to confirm the performance increase after introducing the RAG module.

\mylist{Model of Choice}
We chose GPT-4o (gpt-4o-2024-08-06) \cite{openai_gpt4o}, the newest version of large language model offered by OpenAI at the time of the experiment as our model of choice. The temperature is set to 1.0 during sampling. 

\mylist{Dataset Preparation and Example Pool Creation}
For experiments with \verusbench \cite{yang2024autoverus}, the dataset used for retrieval consists of unverified and verified Rust files from three benchmarks:\footnote{\autoverus was also tested on the unpublished Verus-CloverBench} Diffy, MBPP, and Misc. In addition, example Rust files originally used in the refinement phase of the \autoverus \cite{yang2024autoverus} pipeline were also included. For experiments on \bench, only code documents within the repository are indexed and retrieved.

\mylist{RAG modules}
For few-shot example retrieval, we experimented with both the code index and the informalization index. For dependency retrieval, due to limited amount of data in our current dataset, we only evaluate the simple embedding-based dependency retrieval approach. We leave the more advanced approaches for dependency retrieval such as dependency graphs and embedding projection for future work.

\subsection{Evaluation on Function-Level Verification}

While the full \autoverus pipeline is already highly effective on this relatively small and constrained dataset (proving 137 out of 150 tasks in \verusbench)~\cite{yang2024autoverus}, our goal here is to demonstrate the effectiveness of RAG, which represents an orthogonal contribution. Therefore, we choose our baseline to be the direct-generation pipeline presented in the original \autoverus paper. While the original process samples up to 125 answers, we limit sampling to 5 times for each of our setups to examine the benefits from retrieval modules.

The performance comparison between different retrieval strategies is shown in Table \ref{tab:Verus_bench_result_percent}. \ragverus-Code denotes the results achieved by performing retrieval on the code index; similarly, \ragverus-Text retrieves on the informalization index.

\ragverus-Code consistently outperformed other methods, achieving the highest counts of correct, safe, and successful completions overall (87, 134, 84) and excelling particularly in MBPP. Retrieval using informalized summaries also substantially improved over the Baseline, with a notable edge over code-based retrieval in count of success for Diffy tasks (23 vs. 19). The experiment results confirmed that context retrieval methods contribute to the overall success of \ragverus. Comparing to the baseline results reported in the \autoverus paper, the same baseline could only solve 67 tasks even with a maximum LLM invocation allowance of 125.

\begin{table}[ht!]
\centering
\setlength\extrarowheight{-3pt}
\caption{Evaluation results for \verusbench (MBPP, Diffy, Misc, and All tasks) with percentages based on total number of tasks in each category.}
\label{tab:Verus_bench_result_percent}
\begin{tabular}{llccc}
\toprule
\textbf{Task}  & \textbf{Model}   & \textbf{Correct (n, \%)} & \textbf{Safe (n, \%)} & \textbf{Success (n, \%)} \\
\midrule
 & Baseline        & 23 (29.5\%)         & 60 (76.9\%)      & 17 (21.8\%)         \\
\textbf{MBPP} & \ragverus\ - Code       & 57 (73.1\%)         & 74 (94.9\%)      & 54 (69.2\%)         \\
& \ragverus\ - Text    & 49 (62.8\%)         & 68 (87.2\%)      & 45 (57.7\%)         \\
\midrule
& Baseline        & 3 (7.9\%)           & 30 (78.9\%)      & 2 (5.3\%)           \\
\textbf{Diffy} & \ragverus\ - Code       & 19 (50.0\%)         & 38 (100.0\%)     & 19 (50.0\%)         \\
& \ragverus\ - Text    & 24 (63.2\%)         & 37 (97.4\%)      & 23 (60.5\%)         \\
\midrule
& Baseline        & 7 (30.4\%)          & 21 (91.3\%)      & 6 (26.1\%)          \\
\textbf{Misc} & \ragverus\ - Code       & 11 (47.8\%)         & 22 (95.7\%)      & 11 (47.8\%)         \\
& \ragverus\ - Text    & 8 (34.8\%)          & 22 (95.7\%)      & 8 (34.8\%)          \\
\midrule
& Baseline        & 33 (23.7\%)         & 111 (79.9\%)     & 25 (18.0\%)         \\
\textbf{All} & \ragverus\ - Code       & \textbf{87 (62.6\%)} & 134 (96.4\%)     & \textbf{84 (60.4\%)} \\
& \ragverus\ - Text    & \textbf{81 (58.3\%)} & 127 (91.4\%)     & \textbf{76 (54.7\%)} \\
\bottomrule
\end{tabular}
\end{table}

\subsection{Evaluation on \bench}

Although we still examine proof completion on a per-function basis, repository-level verification becomes more challenging due to the need for coordinated reasoning across interdependent modules.

We run ablated trials of \ragverus on the new \bench for proof-completion; specific model parameters are listed in Appendix \ref{appen:benchSetup}. We compare to two baselines, the direct-generation as well as the refinement pipelines from \autoverus; we then augment each pipeline with our aforementioned retrieval module with code index. Resulting pass rates are shown in Table \ref{tab:Proj_bench_result_percent}, reflecting a remarkable improvement on every setting augmented by context retrieval.

\vspace*{-1.5\baselineskip}
\begin{table}[ht!]
\centering
\setlength\extrarowheight{-1pt}
\caption{Evaluation results on \bench (\verismo-Simple$+$Complex).}
\label{tab:Proj_bench_result_percent}
\begin{tabular}{llccc}
\toprule
\textbf{Task}  & \textbf{Model}   & \textbf{Correct (n, \%)} & \textbf{Safe (n, \%)} & \textbf{Success (n, \%)} \\
\midrule
& DirectGen greedy & 2 (3.8\%) & \textbf{52 (100.0\%)} & 2 (3.8\%) \\
& DirectGen sample & 4 (7.7\%) & 48 (92.3\%) & 4 (7.7\%) \\
\textbf{Simple} & Refinement (\autoverus) & 8 (15.4\%) & 49 (94.2\%) & 7 (13.5\%) \\
52 tasks & DirectRAG & 21 (40.4\%) & \textbf{52 (100.0\%)} & 21 (40.4\%) \\
& \textbf{Refinement+RAG} & \textbf{24 (46.2\%)} & 45 (86.5\%) & \textbf{23 (44.2\%)} \\
\midrule
& Refinement (\autoverus) & 63 (16.4\%) & \textbf{294 (76.8\%)} & 59 (15.4\%) \\
\textbf{Overall} & DirectRAG & 78 (20.4\%) & 281 (73.4\%) & 65 (17.0\%) \\
383 tasks& \textbf{Refinement+RAG} & \textbf{84 (21.9\%)} & 266 (69.5\%) & \textbf{75 (19.6\%)} \\
\bottomrule
\end{tabular}
\end{table}
\vspace*{-\baselineskip}

We note that even in the Complex setting, it is still a simplification over real-world verification challenges, as we assume that proof annotations are only erased for one function at a time. Nevertheless, we observe that information retrieved from the same repository provides constructive in-distribution examples to complete successful proofs, especially in the Simple category.
Yet, over 55\% Simple tasks are not solved by either model.

We argue that \bench-Complex is still a very challenging task; over the actual 331 hard cases, the combination \texttt{Refinement+RAG} (75-23=52) only solves less than 16\% of verification tasks, suggesting further innovations in both retrieval methods and generative agents would be necessary. For interested readers, qualitative analysis of the proof generation is discussed in Appendix~\ref{appen:benchQuali}.

\section{Related Work}

\mylist{Automated program verification with LLM} While there exist various techniques for automated program verification with machine learning based methods like Code2Inv~\cite{si2020code2inv}, CIDER~\cite{liu2022learning} and Code2RelInv~\cite{wang2022learning}, there has been recent advancements focusing on the integration of LLMs to enhance proof generation capabilities~\cite{wen2024enchanting}. LLMs provide the potential to automate these processes by generating human-like proofs~\cite{wen2024enchanting} and code~\cite{li2024guiding}. LLM-aided proof/code generation has been studied within verification-aware programming languages like Frama-C~\cite{kirchner2015frama}, Dafny~\cite{leino2010dafny} and Verus~\cite{lattuada2023verus}. Recent work \autoverus ~\cite{yang2024autoverus} leverages LLMs for automated proof synthesis in Rust using finetuned knowledge bases and refinement processes. Clover~\cite{sun2024clover} addresses the case of SV where no specification is formally given, and attempts to autoformalize a specification in Dafny by aligning with the available implementations and documentations. LeanDojo~\cite{yang2023leandojo} introduces a large premise pool in Lean and uses fine-tuned retrieval models to perform RAG for proofs. Although these approaches differ in methodology and application areas, they all focus primarily on single-function verification. 

\mypara{Repository-level program verification} Repository-level program verification is a relatively emerging field with limited previous research addressing the inherent complexity of large software systems. Selene~\cite{zhang2024selene} represents a pioneering effort in this domain, but the language focus, verification tooling, and dependency management approaches are different from \ragverus. While there has been recent development in repository-level LLM-based code generation~\cite{liao20243}, \ragverus extends this domain into automated verification, ensuring not only the generation of code but also its formal correctness.

\section{Conclusion and Future Work}

\ragverus addresses repository-level verification via retrieval-augmented generation and context-aware prompting, enabling LLMs to synthesize proofs informed by cross-module dependencies and project-wide examples. Supporting this effort, \bench provides the first repository-level benchmark for Verus, derived from real-world systems to reflect compositional reasoning challenges, and establishes a configurable playground for evaluating retrieval strategies. Future directions include implementing more fine-grained retrieval methods and expanding the benchmark to diverse repositories, possibly using self-evolved methods on verification code~\cite{chen2024evolveVerus}.
By bridging AI-driven synthesis with practical project demands, we hope this work lays a foundation for realistic, community-driven assessments on repository-level program verification tools.

\bibliographystyle{splncs04}
\bibliography{ragverus}

\newpage

\appendix

\section*{Appendix}

\section{\bench Data Example}
\label{appen:bench_example}
We process code data to generate tasks and to prepare informative contexts, following the methods outlined in Section 4.1.

\begin{figure*}[ht!]
    \noindent
    \centering
    \centerline{
    \adjustbox{max width=\textwidth}{
        \begin{subfigure}[b]{\textwidth}
            \centering
            \includegraphics[width=\textwidth]{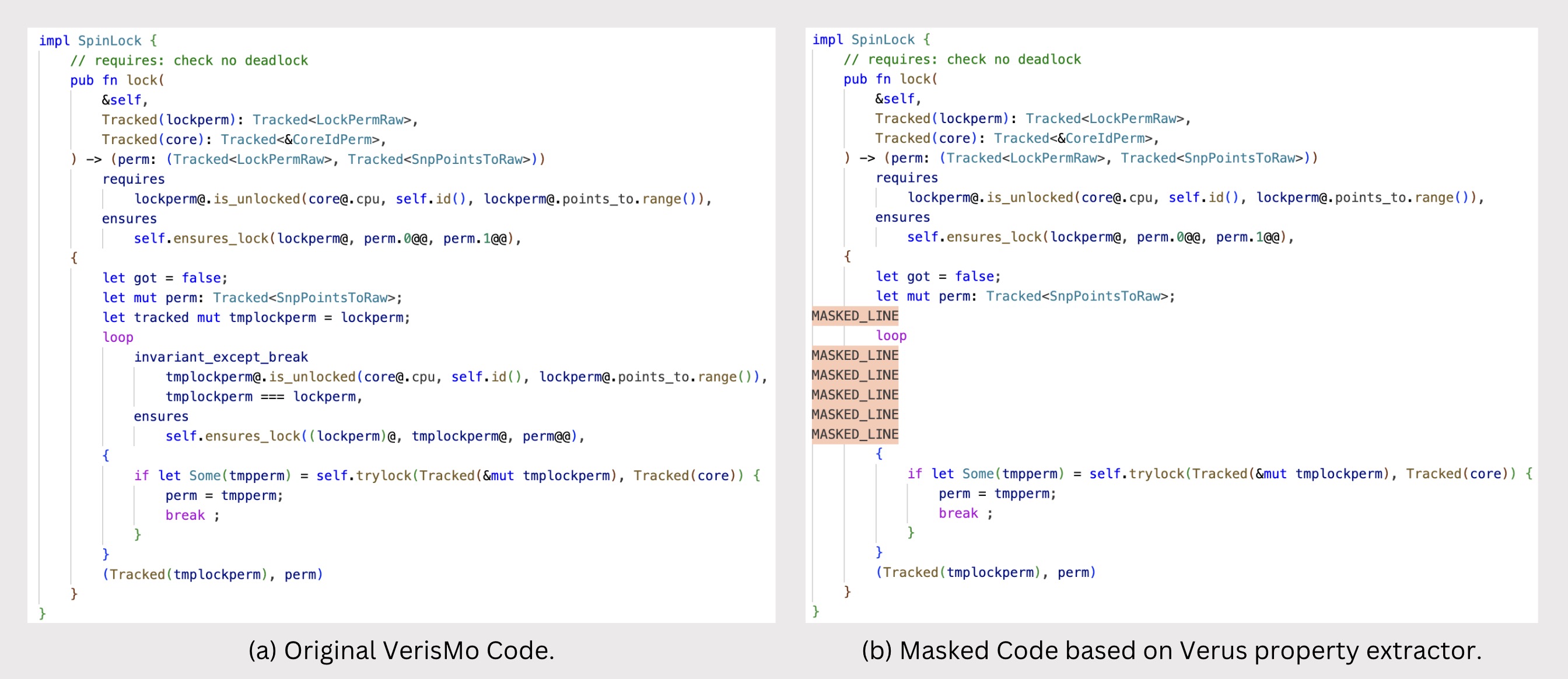}
            \caption{}
            \label{fig:masked_code}
        \end{subfigure}
    }}
    \adjustbox{max width=\textwidth}{
        \begin{subfigure}[b]{0.6\textwidth}
            \centering
            \includegraphics[width=\textwidth]{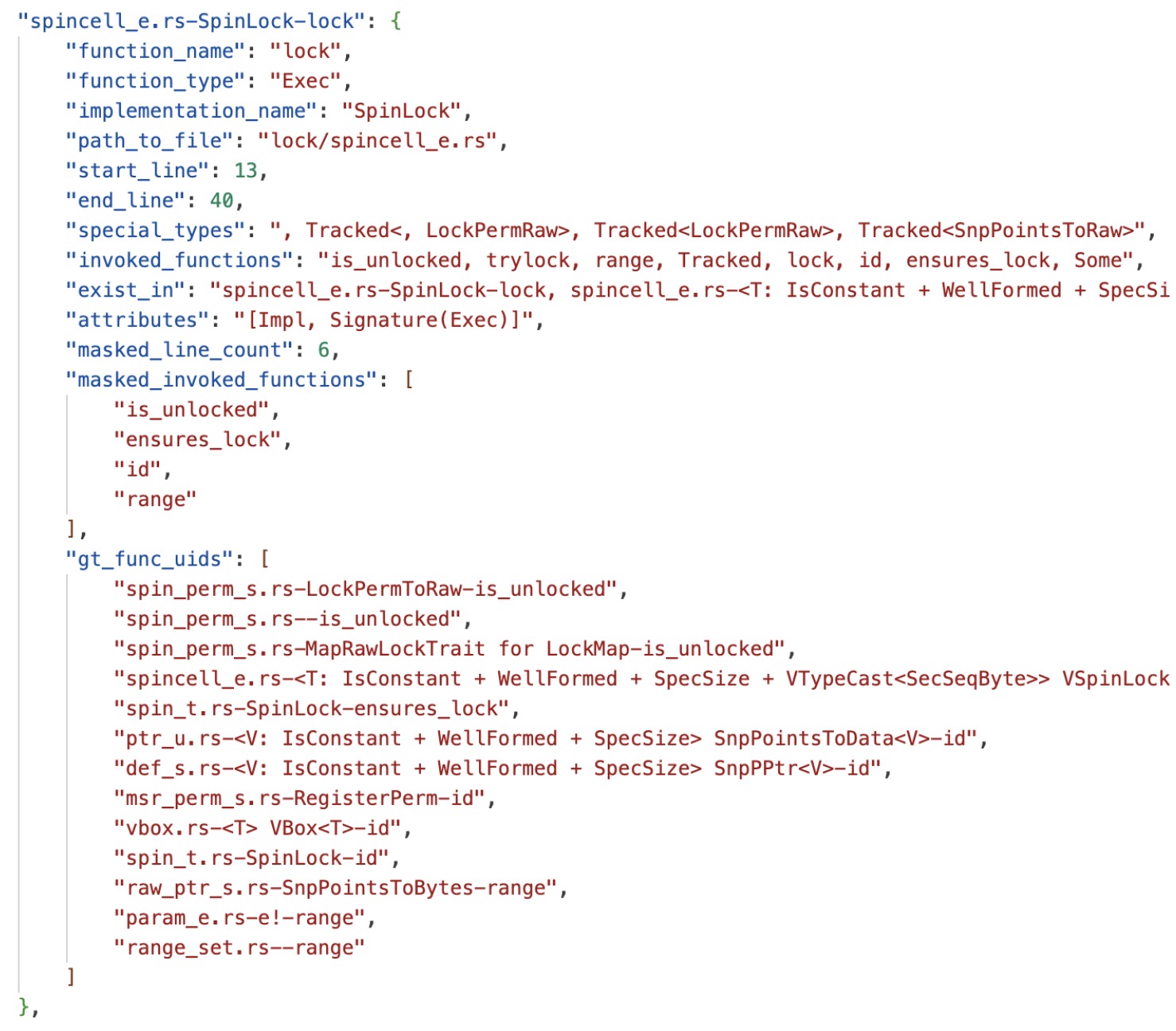}
            \caption{}
            \label{fig:metadata}
        \end{subfigure}
    }
    \caption{Examples of code masking and metadata extraction using Verus-adapted tools. (a) an example of an original function and its masked counterpart, where proof annotations are replaced with \texttt{MASKED\_LINE} (highlighted in orange); (b) an example of the extracted metadata  }
    \label{fig:process_benchmark}
\end{figure*}

\section{\bench Experiment Setup}
\label{appen:benchSetup}

We maintain a similar LLM budget across experiments when evaluating on the \bench in Section 5.2, so as to ensure a fair comparison between different approaches.

For all methods that require sampling, we set the temperature to t=1.0 for diverse generation results. Except for \texttt{DirectGen greedy}, which is our deterministic baseline, we set t=0 and only sample once.

\mylist{Direct Generation}
For \texttt{DirectGen-sample} and \texttt{DirectRAG}, we set the number of generations to a constant of 3 samples.

\mylist{Refinement Generation}
For \texttt{Refinement} and \texttt{Refinement+RAG}, we generate 2 initial samples and allow a maximum of 2 more repair steps, budgeting under 4 LLM calls.

\vspace{5pt}
We note again that the retrieval module used in Section 5.2 searches over \verismo contents only, and only uses code embedding information when conducting similarity retrieval.


\section{\bench Qualitative Experiment Results}
\label{appen:benchQuali}

\mypara{Common failure modes in \bench}

\verismo contains nested dependencies in proofs and often requires special formats of annotation different from common examples given by the official Verus tutorial.

Especially for models without our RAG assistance, we observe on the Simple dataset that they would produce answers that are logically correct to human, but miss subtle syntax in \verismo, such as not using the specially defined integer type for the particular module.


\mypara{Key Difficulties Faced in \bench-Complex}

In Table~\ref{tab:Proj_bench_result_percent}, as reflected by the success rates between \texttt{Refinement} (59-7=52) and \texttt{Refinement+RAG} (75-23=52) on \bench-Complex, the basic context references provided by the current retrieval method are too simple to provide much assistance in completing the hard verification tasks.

The current experiments fail to handle several situations:
\begin{itemize}
    \item We observe that many contextually similar tasks in the hard category require different premise sets and proof style, requiring more specialized direction of retrieval
    \item The ground-truth premise pool is actually larger than the maximum number we return, demanding larger context capacity from the retrieval module
    \item Some tasks require super long proofs (> 80 proof annotation lines). We suspect any successful run would require hundreds of refinement cycles and compiler feedbacks, in addition to a fully complete premise pool, which is out of our current sampling budget
\end{itemize}

\mypara{Generation Style}
Since the number of verification success, as a binary metric, does not capture gradual improvement in code generation quality, We evaluate the code similarity with respect to the ground truth proof to reflect how much of the proofs are on the right directions. We analyze the average BLEU scores for selected pipeline settings from Section 5.2, calculated over the entire \bench dataset.

We observe that both retrieval augmented pipelines produce more coherent answers, suggesting that they are utilizing the \verismo-specific contexts as expected.

\begin{table}[h!]
\centering
\caption{BLEU Scores between Method-Generated Answers and the Ground-Truth}
\label{table:bleu_scores}
\begin{tabular}{>{\bfseries}l c}
\toprule
\textbf{Code Source} & \textbf{Average BLEU Score} \\
\midrule
Unverified Query & 46.76 \\
DirectRAG & \textbf{57.75} \\
Fullpipe Base & 48.18 \\ 
Fullpipe RAG & 55.97 \\
\bottomrule
\end{tabular}
\end{table}

\end{document}